\documentclass{article}
\usepackage{epsf}
\usepackage{color}
\newcommand{\Cont}{\Upsilon} % Contraste
\newcommand{\cj}{{}^*} % conjugue
\renewcommand{\t}{{^{\rm\scriptscriptstyle T}}} % transposition
\newcommand{\pinv}{{}^-} % pseudo-inverse
\newcommand{\qref}[1]{(\ref{#1})}
\newcommand{\krons}{\oslash{}}
\newcommand{\kronc}{\otimes{}}
\newcommand{\out}{\circ}  % outer product
\newcommand{\con}{\mathop{\bullet}} % contraction
\newcommand{\vecs}[1]{{\rm\bf vecs}\{#1\}}
\newcommand{\Unvecs}[2]{{\rm\bf Unvecs}_{#1}\{#2\}}

\newcommand{\vect}[2]{\left(\begin{array}{c} #1\\#2\end{array}\right)}

\newcommand{\eqdef}{\stackrel{\rm def}{=}} 

\newcommand{\Max}{\mathop{\rm Max}}

\newcommand{\Ker}[1]{{\rm Ker}\{#1\}}
\newcommand{\E}{{\rm E}}
\newcommand{\Cum}[1]{{\rm Cum}\{#1\}}

\renewcommand{\j}{\jmath}  % racine de -1
\newcommand{\un}{{\bf 1}}
\newcommand{\zero}{{\bf 0}}
\newcommand{\bm}[1]{\mbox{\boldmath $#1$}} % bold
\newcommand{\bms}[1]{\mbox{\boldmath$\scriptstyle #1$}} % boldPetit
\newcommand{\ba}{\bm{a}}
\newcommand{\bff}{\bm{f}}
\newcommand{\bg}{\bm{g}}
\newcommand{\bi}{\bm{i}}
\newcommand{\bj}{\bm{j}}

\newcommand{\bn}{\bm{n}}
\newcommand{\bt}{\bm{t}}
\newcommand{\bu}{\bm{u}}
\newcommand{\bv}{\bm{v}}
\newcommand{\bw}{\bm{w}}
\newcommand{\bx}{\bm{x}}
\newcommand{\by}{\bm{y}}
\newcommand{\bz}{\bm{z}}
\newcommand{\bA}{\bm{A}}
\newcommand{\bB}{\bm{B}}
\newcommand{\bC}{\bm{C}}
\newcommand{\bF}{\bm{F}}
\newcommand{\bG}{\bm{G}}
\newcommand{\bH}{\bm{H}}
\newcommand{\bI}{\bm{I}}
\newcommand{\bM}{\bm{M}}
\newcommand{\bN}{\bm{N}}
\newcommand{\bP}{\bm{P}}
\newcommand{\bQ}{\bm{Q}}
\newcommand{\bR}{\bm{R}}
\newcommand{\bS}{\bm{S}}
\newcommand{\bT}{\bm{T}}
\newcommand{\bU}{\bm{U}}
\newcommand{\bY}{\bm{Y}}
\newcommand{\bZ}{\bm{Z}}
\newcommand{\bLambda}{\bm{\Lambda}}
\newcommand{\bSigma}{\bm{\Sigma}}

\newcommand{\cB}{\mathcal{B}}
\newcommand{\cC}{\mathcal{C}}

\newcommand{\cM}{\mathcal{M}}

\newcommand{\ie}{{\it i.e.}}
\newcommand{\eg}{{\it e.g.}}
\newcommand{\nbc}{K} % nb capteurs
\newcommand{\nbs}{P} % nb sources
\newcommand{\nbe}{N} % nb echantillons

\newcommand{\scal}[1]{\langle #1 \rangle} % prod scalaire

\newcounter{theorem}
\newtheorem{theorem}{Theorem}[section]
\title{Mathematics in Signal Processing V}
\author{J. G. McWhirter and I. K. Proudler Eds.}
\date{}
\begin{document}\sloppy
\pagenumbering{roman}

\maketitle

\bigskip

\centerline{Oxford University Press}

\vfill

\newpage
\tableofcontents
%\contributorlist
%\item[Pierre Comon] Lab. I3S, CNRS, BP121, F-06903
%Sophia-Antipolis cedex, France\\
%\endcontributorlist
%\chapter[Tensor Decompositions]{Tensor Decompositions\\ \normalsize State of the Art and Applications}
%\author[P.\ Comon]{Pierre Comon}
%\address{CNRS}
\newpage
\pagenumbering{arabic}
%\maketitle
\thispagestyle{empty}
\setcounter{page}{1}

\begin{center}
\huge Tensor Decompositions\\ 
\normalsize State of the Art and Applications

\bigskip\bigskip

{\bf Pierre Comon}\\
{\it Lab. I3S, CNRS, BP121, F-06903
Sophia-Antipolis cedex, France}
\date{}
\end{center}

\begin{abstract}
In this paper, we present a partial survey of the tools borrowed from
tensor algebra, which have been utilized recently in Statistics and
Signal Processing. It is shown why the decompositions well known in
linear algebra can hardly be extended to tensors. The concept of rank is itself difficult to define, and its calculation raises difficulties.
Numerical algorithms have nevertheless been developed, and some are
reported here, but their limitations are emphasized. These reports
hopefully open research perspectives for enterprising readers.
\end{abstract}

%%%%%%%%%%%%%%%%%%
%%%% REPRINTS %%%%
%\pagestyle{plain}\setlength{\footskip}{8mm}%\setcounter{page}{xx}
%\vspace{-11.2cm}\hspace{2.3cm}\mbox{Oxford}\vspace{11.1cm} %%% REPRINTS
% FOR REPRINTS
\vspace{-10cm}\begin{minipage}{0.9\textwidth}\begin{center}
in {\it ~\hspace{-2em} Mathematics in Signal Processing V},
J. G. McWhirter and I. K. Proudler Eds.,
Oxford University Press, Oxford, UK, 2001
\end{center}\end{minipage}\vspace{9.5cm}

%%%%%%%%%%%%%%%%%%%%%
%%%%%%%%%%%%%%%%%%%%%
\section{Introduction}
{\bf Applications. }
The decomposition of arrays of order higher than 2 has proven to be
useful in a number of applications. The most striking case is perhaps
{\em Factor Analysis}, where statisticians early identified difficult
problems, tackling the limits of linear algebra. The difficulty lies in
the fact that such arrays may have more factors than their dimensions.
Next, data are often arranged in many-way arrays, and the reduction to
2-way arrays sometimes results in a loss of information. Lastly, the
solution of some problems, including the so-called {\em Blind Source
Separation} (BSS) generally requires the use of High-Order Statistics
(HOS), which are intrinsically tensor objects \cite{Mccu87} (McCullagh
1987). When second order statistics suffice to establish
identifiability, the corresponding algorithms are quite sensitive to
model uncertainties \cite{LiavRD99:ieeesp} (Liavas Regalia and Delmas
1999), so that the complementary use of HOS statistics is often still
recommended.

BSS finds applications in Sonar, Radar \cite{ChauCM93:ieeF} (Chaumette
Comon and Muller 1993), Electrocardiography \cite{DelaDV00:ieeebe}
(DeLathauwer DeMoor at alterae 2000), Speech \cite{NguyJKC96:asp}
\cite{LeeLGS99:spl} \cite{Dela97:these} (Nguyen-Thi and Jutten 1996;
Lee and Lewicki 1999; DeLathauwer 1997), and Telecommunications
\cite{FerrC00:ieeesp} \cite{GassG97:ieeesp} \cite{VandP96:ieeesp}
\cite{CastM97:spawc} \cite{GrelC00:tampere} (Ferreol and Chevalier 2000;
Gassiat and Gamboa 1997; Van der Veen 1996; Castedo and Macchi 1997;
Grellier and Comon 2000), among others. In particular, the surveillance
of radio-communications in the civil context, or interception and
classification in military applications, resort to BSS. Moreover, in
{\em Mobile Communications}, the mitigation of interfering users and the
compensation for channel fading effects are now devised with the help of
BSS; this is closely related to the general problem of {\em Blind
Deconvolution}.

High-Order Factor Analysis is applied in many areas including Economy,
Psychology \cite{CarrC70:psy} \cite{CaroPK80:psy} (Carroll and Chang
1970; Carroll Pruzansky and Kruskal 1980), Chemometrics
\cite{Gela89:cils} \cite{Bro97:cils} (Geladi et alterae 1989; Bro
1997), and Sensor Array Processing 
\cite{Como98:spie} \cite{VandP96:ieeesp} \cite{SidiBG00:ieeesp} (Comon
1989; Van der Veen and Paulraj 1996; Sidiropoulos Bro and Giannakis
2000; Comon 2000). Other fields where array decompositions can turn out
to be useful include Exploratory Analysis \cite{JoneS87:jrss} (Jones and
Sibson 1987), Complexity Analysis \cite{Krus77:laa}
\cite{Howe78:laa} (Kruskal 1977; Howell 1978), and
Sparse Coding \cite{HyvaHOG99:aussois} (Hyv\"arinen Hoyer and Oja 1999).

{\bf Bibliographical survey. }
Bergman \cite{Berg69:ja} (1969) and Harshman \cite{Hars72:ucla}
(1970) were the first to notice that 
the concept of {\em rank} was difficult to extend from matrices to
higher order arrays. Carroll \cite{CarrC70:psy} (1970) provided the
first {\em canonical decomposition} algorithm of a three-way array,
later referred to as {\em Candecomp} model. Several years later, Kruskal
\cite{Krus77:laa} (1977) conducted a detailed analysis of uniqueness,
and related several definitions of rank. The algorithm {\em Candelinc}
was devised by Carroll and others in the eighties \cite{CaroPK80:psy}
(Carroll et alterae 1980); it allowed to compute a canonical
decomposition subject to a priori linear constraints.

Leurgans and others \cite{LeurRA93:simax} (1993) derived sufficient
identifiability conditions for the 3-way array decomposition; as opposed
to Kruskal, his proof was constructive and yielded a numerical algorithm
running several matrix SVD's.

Instead of finding an exact decomposition of a $d-$way array, which
requires more than $d$ terms (as we shall subsequently see), Comon
proposed \cite{Como91:cha} (1991) to approximately decompose it into
$d$ terms. The problem was then reduced to finding an invertible linear
transform (change of coordinates); see \cite{Como94:SP}
\cite{CardS93:ieeF} (Comon 1994; Cardoso 1993) and references therein.
This decomposition is now referred to as ``{\em Independent Component
Analysis}'' (ICA), whereas the exact {\em Canonical Decomposition} is
sometimes referred to as {\em underdetermined} or {\em over-complete} ICA
\cite{LeeLGS99:spl} \cite{Como98:spie} (Lee et alterae 1999; Comon
1998). 

The terminology of ICA is meaningful in the context of Signal
Processing and BSS \cite{JuttH91:SP} \cite{Como94:SP}
\cite{Card99:nc} (Jutten and H{\'e}rault 1991; Comon 1994; Cardoso 1999).
Constructive algorithms for ICA either proceed by sweeping the pairs of
indices \cite{Como89:vail} \cite{Como94:SP} \cite{CardS93:ieeF}
(Comon 1989; Comon 1991; Cardoso and Souloumiac 1993; Comon 1994), or
are of iterative nature, like power methods \cite{DelaCD95:nolta}
\cite{KofiR00:ams} (DeLathauwer Comon and others 1995; Kofidis and
Regalia 2000), gradient descents \cite{MaccM97:ieeesp} (Macchi and
Moreau 1997), or Robbins-Monro algorithms \cite{JuttH91:SP}
\cite{NguyJKC96:asp} (Jutten and Hérault 1991; Nguyen-Thi Jutten and
others 1996). Some less efficient early methods were based on contracted
versions of the array \cite{Card89:gla} (Cardoso 1989)
or on noiseless observations\cite{Como89:vail} (Comon 1989).

A solid account on decompositions of 3-way arrays can also be found in
DeLathauwer's PhD thesis \cite{Dela97:these} (1997); an interesting
tool defined therein is the {\em HOSVD} \cite{DelaBV93:athos}
(DeLathauwer and others 1993), generalizing the concept of SVD to arrays
of order 3, in a different manner compared to Carroll, but quite similar
to the {\em Tuckals} decomposition \cite{Levi65:pb} \cite{Tuck66:psy}
\cite{Gela89:cils} (Levin 1965; Tucker 1965; Geladi 1989). A good survey of
rank issues can also be found in \cite{KofiR00:ams} (Kofidis et
alterae 2000). An account on identifiability issues can be found in
\cite{CaoL96:ieeesp} (Cao and Liu 1996). 

\nocite{CoppB89}

%%%%%%%%%%%%%%%%%%%%%
%%%%%%%%%%%%%%%%%%%%%
\section{Tensors}
%%%%%%%%%%%%%%%%%%%%%
\subsection{Terminology}
The {\em order} of an array refers to the number of its {\em ways}; the
entries of an array of order $d$ are accessed via $d$ indices, say
$i_1..i_d$, with every index $i_a$ ranging from 1 to $n_a$. The integer $n_a$
is one of the $d$ {\em dimensions} of the array. For instance, a matrix
is a $2-$way array (order 2), and thus has 2 dimensions. A vector is an
array of order 1, and a scalar is of order 0.

Throughout this paper, and unless otherwise specified, variables take
their values in the real field, 
although all the statements hold true in the complex field with more
complicated notations; boldface lowercase letters, like $\bu$, will
denote single-way arrays, \ie\ vectors, whereas boldface uppercase
letters, like $\bG$, will denote arrays with more than one way, \ie\
matrices or many-way arrays. The entries of arrays are scalar quantities
and are denoted with plain letters, such as $u_i$ or $G_{ijk\ell}$.

A tensor of order $d$ is a $d-$way array that enjoys the {\em multilinearity
property} after a change of coordinate system. For instance,
consider a $3$rd order tensor $\bT$ with entries $T_{ijk}$, and a
change of coordinates defined by 3 square invertible matrices, $\bA$, $\bB$
and $\bC$. Then, in the
new coordinate system, the tensor $\bT'$ can be written as a function of
tensor $\bT$ as:
\begin{equation}\label{multilin:eq}
T'_{ijk} = \sum_{abc} A_{ia} B_{jb} C_{kc} T_{abc}
\end{equation}
In particular, moments and cumulants of random variables may be treated
as tensors \cite{Mccu87} (McCullagh 1987). This product is sometimes
referred to as the 
Tucker product \cite{KofiR00:ams} (Kofidis et al. 2000) between
matrices $\bA$, $\bB$, and 
$\bC$, weighted by $\bT$. Note that tensors enjoy property
\qref{multilin:eq} even if the above matrices are not invertible; only
linearity is required.

Tensor algebra is a well identified framework; in particular, two kinds
of indices are distinguished, covariant or contravariant, depending on
the role they play in the application under consideration: an array can
indeed be seen as an {\em operator} from one space to another. For the
sake of simplicity, we {\em shall not} pay too much attention to this 
distinction, although it turns out to be
important in contexts other than the present one. 

\subsection{Notation}\label{notation:sec}
Given two arrays of order $m$ and $n$, one defines their {\em outer
product} $\bC=\bA \out \bB$ as the array of order $m+n$:
\begin{equation}
C_{ij..\ell\,ab..d} = A_{ij..\ell} \, B_{ab..d}
\end{equation}
For instance, the outer product of two vectors, $\bu\out\bv$, is a matrix.

\smallskip

Given two arrays, $\bA=\{A_{ij..\ell}\}$ and $\bB=\{B_{i'j'..\ell'}\}$
of orders $d_A$ and $d_B$ respectively, having the same first dimension,
one can define the {\em mode$-1$} contraction product:
$$
\left(A \con B \right)_{j..\ell j'..\ell'} = \sum_{i=1}^{n_1}
A_{ij..\ell}B_{ij'..\ell'} 
$$
For instance, the standard matrix-vector product is $\bA\,\bu =
\bA\t\con\bu$. 
Similarly, one defines the {\em mode$-p$} inner product when arrays
$\bA$ and $\bB$ have the same $p$th dimension, by summing over the $p$th
index; the product is denoted as 
$$
\bA \con_p \bB
$$
If unspecified, the contraction applies by default to the first index.
Some authors denote this product as $\bA {\times}_p \bB$, but we find it
less readable.

We define the Kronecker product $\bu\kronc\bv$ between two vectors $\bu$
and $\bv$ as the vector containing all the possible cross-products
\cite{Brew78:cs} (Brewer 1978). If $\bu$ and $\bv$ are of dimension
$J$ and $K$, then $\bu\kronc\bv$ is of dimension $JK$. 

Lastly, we define the symmetric Kronecker product of a $\nbc-$ dimensional
vector $\bw$ by
itself, denoted $\bw\krons\bw=\bw^{\krons2}$, as the
$\nbc((\nbc+1)/2-$dimensional vector containing all the distinct
products, with an appropriate weighting of the cross terms so that
$||\bw\krons\bw|| = ||\bw\otimes\bw||$. The product $\bw^{\krons d}$ is
defined in a similar manner for $d>2$. For instance, $\bw^{\krons3}$ is
of size $K(K+1)(K+2)/6$.

The $\vecs{\cdot}$ operator puts a $\nbc\times\nbc$ symmetric matrix in
the form of a $\nbc(\nbc+1)/2-$dimensional vector; conversely,
$\Unvecs{}{\cdot}$ puts 
it back in matrix form. For instance, 
$\Unvecs{}{\bff^{\krons 2}}=\bff\,\bff\t$.
Refer to \cite{GrelC00:tampere} \cite{VandP96:ieeesp} (Van der Veen
1996; Grellier and Comon 2000) for a more detailed description.

%%%%%%%%%%%%%%%%%%%%%
\subsection{Homogeneous polynomials}
$d-$way arrays can be written in two different manners, as pointed out
by several authors \cite{Mccu87} (McCullagh 1987), related to each
other by a bijective mapping, $\bff$. Assume the notations
$\bx^{\bms{j}} \eqdef \prod_{k=1}^{\nbc} x_k^{j_k}$ and $|\bj| \eqdef
\sum_k j_k$. Then for homogeneous monomials of degree $d$,
$\bx^{\bms{j}}$, we have $|\bj|=d$.

To start with, take the example of $(\nbc,d)=(4,3)$: one can associate
every entry $T_{ijk}$ to a monomial $T_{ijk}\:x_ix_jx_k$. For instance,
$T_{114}$ is associated with $T_{114}\,x_1^2x_4$, and thus to
$T_{114}\,\bx^{[2,0,0,1]}$; this means that $f([1,1,4])=[2,0,0,1]$.

More generally, the $d-$dimensional vector index $\bi=[i,j,k]$ can
be associated with a $\nbc-$dimensional vector index $\bff(\bi)$
containing the number of times each variable $x_k$ appears in the
associated monomial. Whereas the $d$ entries of $\bi$ take their values
in $\{1,\dots, \nbc\}$, the $\nbc$ entries of $\bff(\bi)$ take their
values in $\{1,\dots, d\}$ with the constraint that $\sum_k f_k(\bi)=d,
\forall \bi$. 

As a consequence, the linear space of symmetric tensors can be
bijectively associated with the linear space of homogeneous polynomials.
To see this, it suffices to associate every polynomial $p(\bx)$ with the
symmetric tensor $\bG$ as:
\begin{equation} 
p(\bx) = \sum_{|\bms{f}(\bms{i})|=d} G_{\bms{i}} \, \bx^{\bms{f}(\bms{i})}
\end{equation}
where $G_{\bms{i}}$ are the entries of $\bG$. The dimension of these
spaces is $S=(^{\nbc+d-1}_{~~~d})$, and one can choose as a basis the set of
monomials: $\cB(\nbc;d)= \{\bx^{\bms{j}}, \, |\bj|=d\}$.

\medskip

\begin{quote}
{\em Example.} Let $p$ and $q$ be two homogeneous polynomials in
$\nbc$ variables, associated with tensors
$\bP$ and $\bQ$, possibly of different orders. Then, polynomial $pq$ is
associated with $\bP\out\bQ$:
$$
p(\bx) q(\bx)=\sum_{\bms{i}}\sum_{\bms{j}} P_{\bms{i}} Q_{\bms{j}}
\bx^{\bms{f}(\bms{i})+\bms{f}(\bms{j})} = \sum_{[\bms{i} \, \bms{j}]}
[\bP\out\bQ]_{[\bms{i} \, \bms{j}]} \, \bx^{\bms{f}({[\bms{i} \, \bms{j}]})}
$$
\end{quote}

\medskip

In practice, it is convenient to take into account the symmetry of the
tensor, by 
defining $c(\bj)$ as the number of times the entry
$T_{\bms{f}^{-1}(\bms{j})}$ appears 
in the array $\bT$: $c(\bj)=|\bj|!/(\bj)!$, where $(\bj)! \eqdef \prod_k
j_k!$. For binary quantics,
$\nbc=2$ and $c(\bj)=(^{\,d}_{j_1}) = (^{\,d}_{j_2})$;
for instance, for $d=4$, $c([3,1])=4$ and $c([2,2])=6$.

Coefficients $\gamma(\bj,p)$ of a polynomial $p$ in basis $\cB$ are
chosen so as satisfy the relation:
$$
p(\bx) = \sum_{|\bms{j}|=d} \gamma(\bj,p) \, c(\bj) \, \bx^{\bms{j}}
$$
Now, both spaces can be provided with a scalar product.
For $d-$way arrays of dimension $\nbc$,
define the Froebenius scalar product:
$$
\scal{\bG,\bH} = \sum_{\bms{i}} \bG_{\bms{i}} \: \bH_{\bms{i}}
$$
and the induced Euclidian norm. 
For homogeneous polynomials of degree
$d$ in $\nbc$ variables, define the scalar product as
$$
\scal{p,q} = \sum_{|\bms{j}|=d}c(\bj)\, \gamma(\bj,p)\, \gamma(\bj,q)
$$
In particular, monomials in $\cB$ satisfy $||\bx^{\bms{j}}||^2=(\bj)!/d!
= 1/c(\bj)$. 
In the case of binary quantics ($\nbc=2$), this was called the {\em
apolar} scalar product \cite{KunR84:ams} (Kung and Rota 1984). The
latter definition has several advantages \cite{ComoM96:SP} (Comon and
Mourrain 1996). For instance, if $a(\bx)$ is a homogeneous linear form with
coefficient vector $\ba$, then
the scalar product $\scal{a^d,q}$ with any homogeneous polynomial
$q$ of degree $d$ turns out to be the value of $q$ at $\ba$:
$$
\scal{a(\bx)^d,q(\bx)} = \sum_{\bms{i}} c(\bi)\, \gamma(\bi,q)\,
\ba^{\bms{i}} = q(\ba)
$$

\bigskip

The interest in establishing a link between tensors and polynomials lies
in the fact that polynomials have been studied rather deeply during the
last century \cite{Salm85} (Salmon 1885). Some of the results obtained will
be useful in this chapter, and in particular the classification of cubics.

%%%%%%%%%%%%%%%%%%%%%
\subsection{Genericity}
A property will be referred to as {\em generic} if it is true on a dense
algebraic subset. The topology used is the standard one for homogeneous
polynomials, namely that of Zariski \cite{Shaf77} (Shafarevich 1977).
Recall that in this topology, the closed subsets are defined by algebraic
equations of the form $p(\bx)=0$, where $p$ is a polynomial. Its
particularity is that two open non empty subsets always intersect; in
other words, the topology is not separated.

\begin{figure}[ht]
\centerline{
\mbox{\epsfxsize=0.2\textwidth
    \epsfbox{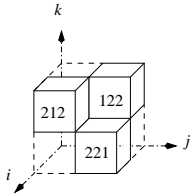}}}
\caption{Non generic example of a binary cubic of maximal rank: position
of non-zero entries in the $2\times2\times2$ associated symmetric tensor.}\label{tensor222:fig}
\end{figure}

For instance, a symmetric matrix is generically of full rank. In fact,
the set of singular matrices is defined by the polynomial relation
$\det(\bA)=0$, which is associated with a closed subset, whose
complementary is dense. We shall subsequently see that this does not
hold true anymore for tensors of order higher than 2. For instance,
the $2\times2\times2$ tensor $\bT$ such that
$T_{122}=T_{212}=T_{221}=1$, and zero elsewhere (cf. figure
\ref{tensor222:fig}), is known to be of rank 3. However, the generic
rank is 2 in that case, as will be discussed in section 
\ref{cand:sec}.

%%%%%%%%%%%%%%%%%%%%%
\subsection{Array ranks}
Let $\bT$ be a tensor, not necessarily symmetric, of dimensions
$n_1\times\dots\times n_d$.  One defines the {\em tensor rank} $\omega$ of
$\bT$ as the minimal number of rank one tensors whose linear
combination yields $\bT$. The properties of tensor rank will be 
extensively discussed in section \ref{cand:sec}.
For completeness, let us also mention the definition of {\em mode$-n$ ranks}.  

The
mode$-n$ vectors of $\bT$ are obtained by varying index $i_n$ and
keeping the others fixed; there are thus as many mode$-n$ vectors as
possibilities of fixing indices $i_k$, $k\neq n$. The mode$-n$ rank,
$R_n$, is defined as the dimension of the linear space spanned by all
mode$-n$ vectors of $\bT$.

\paragraph{Bounds. }
Howell (1978) \cite{Howe78:laa} showed that the tensor rank can be bounded as 
$\omega\leq \max_{i\neq j}\{n_in_j\}$. On the other hand, mode$-n$ ranks and 
tensor rank are related by the inequality $R_n\leq\omega$, $\forall k$.

In the symmetric case, Reznick showed that the tensor rank can be bounded 
as a function of the dimension $\nbc$ and the order, $d$:
\begin{equation}\label{reznick:eq}
\omega \leq \left(\begin{array}{c} \nbc+d-2\\d-1\end{array}\right)
\end{equation}
but this bound is rather loose, except in some very particular cases, as 
will be commented in section \ref{cand:sec}.

%%%%%%%%%%%%%%%%%%%%%
%%%%%%%%%%%%%%%%%%%%%
\section{Cumulants}
\subsection{Definitions}\label{cumdef:sec}
Let $\bz$ be a random variable of dimension $\nbc$, with components
$z_i$. Then one defines its moment and cumulant tensors of order $d$ as:
\begin{eqnarray*}
\cM^{\bms{z}}_{i_1 i_2 .. i_d} &=& \E\{z_{i_1} z_{i_2} \dots z_{i_d}\}\\
\cC^{\bms{z}}_{i_1 i_2 .. i_d} &=& \Cum{z_{i_1}, z_{i_2}, \dots z_{i_d}}
\end{eqnarray*}
When the moment tensors of order less than or equal to $d$ exist and are
finite, the cumulant tensor of order $d$ exists and is finite. Whereas
moments are the coefficients of the expansion of the first
characteristic function $\Phi^{\bms{z}}(\bu)=\E\{exp(\j \bu\t \bz)\}$ about
the origin, where the dotless $\j$ denotes $\sqrt{-1}$, cumulants are
those of the second 
characteristic function, $\Psi^{\bms{z}}(\bu)=\log(\Phi^{\bz}(\bu))$;
for complex random variables, it suffices to consider the joint
distribution of their real and imaginary parts.
Moments and cumulants enjoy the multilinearity property
\qref{multilin:eq} and may be considered as tensors \cite{Mccu87}
(McCullagh 1987).

\smallskip

One important property of cumulant tensors is the following: if at least
two variables, or groups of variables, among $\{z_1, .. z_{\nbc}\}$ are
statistically independent, then all cumulants involving these variables
are null. For instance, if all the
$z_i$ are mutually 
independent, then $\cC^{\bms{z}}_{ij..\ell}=\delta(i,j,..\ell)\,
\cC^{\bms{z}}_{ii..i}$ \cite{KendS77} (Kendall and Stuart 1977), where
the Kronecker $\delta$ is null unless {\em all} its arguments are equal.
This property is not enjoyed by moments, hence the
interest in cumulants.

\smallskip

{\bf The reverse is not true}. In fact, unless the random variable $\bz$ is
Gaussian, an infinite number of cumulants must vanish in order to ensure
their strict sense independence. Therefore, when a cumulant tensor
$\cC^{\bms{z}}$ of order $d$ is diagonal, we shall say that the random
variables $z_i$ are {\em independent at order $d$}.

Gaussian variables play a particular role, since they are the only
random variables that have a finite set of non-zero cumulants
\cite{KagaLR73} \cite{Fell68} (Kagan et al. 1973; Feller 1968). The
cumulants of order 1 and 2 are 
better known under the names of statistical {\em mean} and {\em covariance}.
To illustrate this property,
I looked for a long time for a simple non Gaussian random variable having null
cumulants of order 3 and 4. Take a random variable with values in the
complex plane. If its distribution is 0 with probability one half, and
uniformly distributed on the unit circle with probability one half, then
its mean $\E\{z\}$ is zero, its variance $\E\{zz\cj\}=1$, its second-order
non-circular cumulant $\E\{z^2\}=0$,
its 2 marginal cumulants of order 3, $\E\{z^3\}$ and $\E\{z^2z\cj\}$, are
zero, as well as its 3 marginal cumulants of
order 4, $\Cum{z,z,z,z}$, $\Cum{z,z,z,z\cj}$, $\Cum{z,z,z\cj,z\cj}$.
Yet, this variable is obviously not Gaussian.
This is a very striking example, that one encounters in secondary radar
applications. 

%%%%%%%%%%%%
\subsection{Blind Source Separation}
Consider the linear statistical model
\begin{equation}\label{model:eq}
\by = \bA \, \bx + \bv
\end{equation}
where $\by$ is an observed random variable of dimension $\nbc$, $\bx$ is
a random vector of dimension $\nbs$ referred to as the {\em source
vector}, $\bA$ is a mixing matrix, and $\bv$ stands for background
noise, possible interferers, and measurement errors, independent of
$\bx$.

The {\em Blind Source Separation} (BSS) problem consists of estimating
the mixing matrix, $\bA$, and possibly the corresponding estimates of
$\bx$, solely from measurements of $\by$. In the classical BSS
framework, the components $x_i$ of $\bx$ are assumed to be statistically
independent (generally not in the strict sense because a weaker
independence is sufficient) \cite{Como94:SP} (Comon 1994). In some
cases however, sources $x_i$ may be correlated \cite{VandP96:ieeesp}
\cite{GrelC00:tampere} \cite{ComoG99:aussois} (Van der Veen 1996;
Grellier and Comon 2000; Comon and Grellier 1999), as we shall
subsequently see, and this is not necessarily an obstacle to their
separation. 

To fix the ideas and simplify the notation, assume sources $x_i$ are
independent at order 4. Then, from the properties of cumulants we just
described, we have:
\begin{equation}\label{diago:eq}
\cC^{\bms{y}}_{ijk\ell} = \sum_{p=1}^\nbs A_{ip} A_{jp} A_{kp} A_{\ell p}
\cC^{\bms{x}}_{pppp} 
\end{equation}
up to an additive noise term, $\cC^{\bms{v}}$.
From measurements of $\by$, it is possible to estimate the cumulant
tensor $\cC^{\bms{y}}$. Estimating $\bA$ then amounts to finding the
decomposition \qref{diago:eq}. In practice, because of the noise $\bv$, this
decomposition is not exact. In addition, since the only property
utilized is the source independence at a given order, matrix $\bA$
can only be identified up to a multiplicative factor $\bP \bLambda$,
where $\bP$ is a permutation and $\bLambda$ is diagonal invertible; see
identifiability issues in \cite{Como94:SP} \cite{CaoL96:ieeesp} (Cao
and Liu 1996; Comon 1994).

{\em Blind Deconvolution} is related to the above BSS modeling in two
respects. First, a convolution with a finite impulse response can always
be written as the product with a T\"oplitz matrix, which means that the
modeling \qref{model:eq} still holds valid, provided matrix $\bA$ is subject
to the T\"oplitz structure \cite{Como94:SP}
\cite{GrigR98:siam} (Comon 1994; Grigorascu and Regalia 1998). Second,
if the source process is linear, then 
extracting the sources is equivalent to computing the 
linear prediction residue \cite{Como94:ifac} (Comon 1994). Then, the
problem reduces 
to an unstructured static separation as in \qref{model:eq}.

%%%%%%%%%%%%%%%%%%%%%
%%%%%%%%%%%%%%%%%%%%%
\section{Array decompositions}
%%%%%%%%%%%%%%%%%%%%%
\subsection{Diagonalization by change of coordinates}\label{diag:sec}
\paragraph{Preprocessing for square mixtures.}
If the mixing matrix $\bA$ is square and invertible, which means that
the number of sources, $\nbs$, is equal to the observation dimension,
$\nbc$, then the BSS problem may be seen as a bijective congruent
transformation (ICA).

Denote $\bR_y$ the covariance matrix of the observation. The goal is to
find an estimate $\bz$ of $\bx$ such that its components $z_i$ are
statistically independent. The first idea is thus to build a vector
$\tilde{\by}=\bT\,\by$ that has a diagonal covariance, yielding
decorrelated components. This can be easily done by searching for a (non
unique) square root factor of $\bR_y$; it can be obtained by a Cholesky
factorization or by an Eigen Value decomposition (EVD) of $\bR_y$. We
then define $\bT$ as the inverse of this factor, so that $\bT\bR_y\bT\t=\bI$. 

With this preprocessing $\bT$, the obtained random variable
$\tilde{\by}$ has a covariance equal to identity. We say that this
variable is {\em standardized}.

Now, it may be more appropriate, when the noise covariance, $\bR_v$, (or
conversely the signal covariance $\bR_s=\bR_y-\bR_v$) is known, 
to build $\bT$ as the inverse of a square root of the signal covariance:
$\bT\bR_s\bT\t=\bI$. In fact, this yields an unbiased solution in the
presence of noise. Unfortunately, neither $\bR_v$ nor $\bR_s$ are known
in general, hence the former procedure based on $\bR_y$.

\smallskip

\paragraph{Preprocessing for rectangular mixtures.}
In practice, one can always reduce the problem to the latter when the
number of sources, $\nbs$, is smaller than the observation dimension,
$\nbc$, in the absence of noise, or when the noise covariance is known.
This is now explained below.

If noise is present, denote $\bT_v$ the inverse of a square root of
$\bR_v$, such that we have $\bT_v\bR_v\bT_v\t=\sigma_v\bI$; if noise is
absent, set $\bT_v=\bI$ and $\sigma_v=0$. Now consider the matrix
$\bT_v\bR_y\bT_v = \bT_v\bR_s\bT_v + \sigma\bI$. Its EVD allows to
detect the number of non-zero eigenvalues in $\bR_s$
\cite{BienK83:ieeesp} (Bienvenu and Kopp 1983), equal to $\nbs$ by
definition, as well as to estimate the source space spanned by the
associated eigenvectors: $\bT_v\bR_y\bT_v = \bU \bSigma \bU\t +
\sigma\bI$. The matrix $\bU$ is here of dimension $\nbc\times\nbs$ and
of full rank. The preprocessing defined as $\bT=\bU\t
\bT_v$ eventually yields a $\nbs-$dimensional standardized vector
$\tilde{\by}=\bT\by$ whose noiseless part has a unit covariance, as in
the previous paragraph.

\smallskip

Lastly, if the mixture is rectangular, but with more sources than
sensors, \ie, $\nbs>\nbc$, the mixture cannot be linearly inverted. Such
mixtures are referred to as {\em underdetermined} or {\em over-complete}, as
already pointed out in the bibliographical survey, and their
identification will be addressed separately in section \ref{cand:sec}.
In such a case, the preprocessing is unuseful, and not recommended.

\paragraph{Orthogonal change of coordinates.}
In the preprocessing, we have done only part of the job. In fact, we
have constructed a matrix $\bT$ such that ideally
$\bT\bA\bA\t\bT\t=\bI$, but this only implies that $\bT\bA = \bQ\t$, for
some $\nbs\times\nbs$ orthogonal matrix $\bQ$. This $\bQ$ factor still remains
undetermined. It is thus necessary to resort statistics of order higher
than 2, namely 3 or 4, unless other hypotheses can be assumed. The
choice between these two possibilities depends on the conditioning of
the problem, directly linked to the value of the diagonal tensor
$\cC^{\bms{x}}$. At order 3, this tensor vanishes for all symmetrically
distributed sources, which strongly limits its use. At order 4, this
tensor is generally non zero, except in some exceptional pathological
cases, as that mentioned in section \ref{cumdef:sec}.

In order to find $\bQ$, one can attempt
to diagonalize (approximately) the cumulant tensor of $\bz=\bQ\,\tilde{\by}$,
$\cC^{\bms{z}}_{ijk\ell}= \sum_{pqrs} Q_{ip} Q_{jq} Q_{kr} Q_{\ell s}
\cC^{\tilde{\bms{y}}}_{pqrs}$. The random variable $\bz$ is eventually an
estimate of the source vector $\bx$; in the absence of noise, we have
$\bz=\bP\bLambda\bx$. Because $\bQ$ is orthogonal, minimizing  the non
diagonal entries is equivalent to maximizing the diagonal ones
\cite{Como91:cha} (Comon 1991), so that $\bQ$ can be determined by
\begin{equation}\label{maxCont:eq}
\bQ = Arg \Max_{\bms{Q}} \Cont_{\alpha,4}\,; \: 
\Cont_{\alpha,4}=\sum_i |\cC_{iiii}^{\bms{z}}|^\alpha
\end{equation}
where $\alpha\geq1$. Several optimization criteria of this type, called
{\em contrasts}, have been proposed \cite{Como94:SP}
\cite{MoreP97:spl} \cite{Dela97:these} \cite{Card99:nc}
\cite{Como01:agadir} (Comon 1994; Moreau and Pesquet 1997; DeLathauwer
1997; Cardoso 1999; Comon 2001) and are justified by Information Theory
arguments. Contrary to the matrix case \cite{GoluV89} (Golub and Van
Loan 1989), it is generally impossible to exactly null the non diagonal
entries of a symmetric tensor of order higher than 2, by just rotating
the coordinate axes. In other words, the class of decompositions
presented in this section lead to rank$-\nbc$ {\em approximations} of
$\nbc-$dimensional symmetric tensors. More will be said in the next
section. Numerical 
ICA algorithms are surveyed in section \ref{algo:sec}.

%%%%%%%%%%%%%%%%%%%%%
\subsection{Decomposition into a sum of rank$-1$ arrays}\label{cand:sec}
When the number of sources, $\nbs$, is strictly larger than the
observation dimension $\nbc$, the previous approach does not apply. In
fact, the matrix $\bA$ now has fewer rows than columns, and the
noiseless relation $\by = \bA\,\bx$ cannot be linearly inverted. In
other words, $\bA$ must be identified without attempting to extract the
sources $x_p$. A symmetric tensor of order $d$ can be expressed via a
Canonical Decomposition (CAND) of the form:
\begin{equation}\label{cand:eq}
\cC^{\bms{y}} = \sum_{p=1}^\omega \gamma(p) \,
\ba(p)\out\ba(p)\out\ba(p)\out\ba(p)
\end{equation}
The number of terms, $\omega$, reaches a minimum when it equals the {\em
tensor rank}. This CAND decomposition allows the identification of
matrix $\bA$ if: (i)~it is unique up to $\bLambda\bP-$indeterminations,
and (ii)~the tensor rank $\omega$ is larger than or equal to the number
of sources, $\nbs$.

\paragraph{Generic rank.}
We report in the tables below the generic value of the {\em tensor rank}
as a function of the dimension $\nbc$ and the order $d$
\cite{ComoM96:SP} (Comon and Mourrain 1996). We also report the
dimension $D$ of the manifold of solutions; when it is zero, it means
that there are a finite number of CAND (at most $d^S$), and there is a 
chance of identifying matrix $\bA$ this way.

\begin{quote}
{\em Example. }
Fore matrices ($d=2$), it is known that a quadratic form cannot be uniquely decomposed into a sum of squares. The manifold of solutions is of dimension $D=\nbc(\nbc-1)/2)$.
\end{quote}

\begin{table}[!ht]
\begin{center}\begin{tabular}{|c|c||c|c|c|c|c|c|c|}\cline{2-9}
\multicolumn{1}{c|}{$\omega$}
 &$\nbc$&{\bf2}&{\bf3}&{\bf4}&{\bf5}&{\bf6}&{\bf7}&{\bf8} \\  
\hline\hline
 ~                  & {\bf 3} & 2  & 4  &  5 &  8 & 10 & 12 & 15 \\
\cline{2-9}
\raisebox{1ex}[1ex][0.1ex]{$d$}
                    &{\bf 4}& 3  & 6  & 10 & 15 & 22 & 30 & 42 \\
\hline \end{tabular}\end{center}
\caption{Generic rank $\omega$ of symmetric tensors as a function of the
dimension $\nbc$ and the order $d$}\label{rank:table}
\end{table}

\begin{table}[!ht]
\begin{center}\begin{tabular}{|c|c||c|c|c|c|c|c|c|}\cline{2-9}
\multicolumn{1}{c|}{$D$}
 &$\nbc$&{\bf2}&{\bf3}&{\bf4}&{\bf5}&{\bf6}&{\bf7}&{\bf8} \\  
\hline\hline
 ~                  & {\bf 3} & 0  & 2  &  0 &  5 & 4 & 0 & 0 \\
\cline{2-9}
\raisebox{1ex}[1ex][0.1ex]{$d$}
                    &{\bf 4}& 1  & 3  & 5 & 5 & 6 & 0 & 6 \\
\hline \end{tabular}\end{center}
\caption{Generic dimension $D$ of the manifold of solutions}\label{dimD:table}
\end{table}

The first striking fact that appears in table \ref{rank:table} is that {\em the rank
can exceed the dimension}, which is not true for matrices. For instance,
it can be seen that $\nbs=5$ 
sources can be identified in dimension $\nbc=4$ with a 3rd 
order cumulant tensor, whereas this number increases to $\nbs=10$ with a
4th order tensor.

One can also deduce from table \ref{dimD:table} that $3$rd order tensors
have a finite number of CAND for even dimensions. For $4$th order
tensors, this is satisfied for dimension 7, but not for lower ones. This
is unfortunate, for $4$th order cumulants are very often better
conditioned than $3$rd order ones. Furthermore, most of the proofs leading to
these tables are not constructive. The only known constructive result is
given by the Sylvester theorem (section \ref{sec:sylvester}).

\begin{table}[ht]
\begin{center}\begin{tabular}{l|l}
\hline
$\mathcal{GI}-$orbit & $\omega(p)$\\
\hline\hline
$x^3$ & 1\\
{\color{blue}$x^3+y^3$} & 2 ~({\bf generic})\\
$x^2 y$ & 3\\
\hline
\end{tabular}\end{center}
\caption{Equivalence classes of binary cubics: orbits under the action
of $\mathcal{GI}$, 
the group of invertible $2-$dimensional changes of
coordinates.}\label{Bincubics:table} 
\end{table}

\paragraph{Non generic rank.} In addition, these results are only valid
in generic cases. And it turns out 
that, contrary to matrices (\ie, 2nd order tensors), {\em the generic rank is
not always maximal}. In other words, the rank can exceed its generic
value. Unfortunately, the maximal achievable rank is not known for all
pairs $(\nbs,\nbc)$. But we can still illustrate this odd fact with particular
values. 

\begin{figure}[ht]
\centerline{
\mbox{\epsfxsize=0.6\textwidth
    \epsfbox{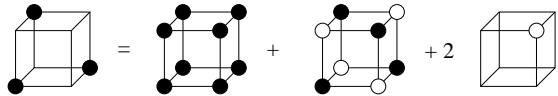}}}
\caption{Explicit decomposition of the non generic example of binary
cubic of maximal rank. Black bullets represent $+1$'s and white bullets
$-1$'s.}\label{decomp333:fig} 
\end{figure}

\begin{quote}
{\em Example.} For instance, for $\nbc=2$ and $d=3$, the maximal rank is 3.
The symmetric tensors having rank 3 are associated with polynomials in
the orbit of $x^2y$. The tensor associated with the latter homogeneous
polynomial is represented in figure \ref{tensor222:fig}, where only 3
entries are equal to 1, the others being null. As reported in table
\ref{Bincubics:table} there is a single class associated with every
value of the tensor rank.

Now to make it more explicit, the polynomial $x^2y$ can be written as:
$$
6\, x^2 y = (x+y)^3 + (-x+y)^3 - 2 y^3
$$
This relation can be rewritten in tensor form as:
$$
\bT = \vect{1}{1}^{\out3} + \vect{-1}{1}^{\out3} - 2 \, \vect{0}{1}^{\out3}
$$
which is an explicit irreducible CAND. This decomposition is depicted in
figure \ref{decomp333:fig}.
Also note that in this case, the Reznick bound \qref{reznick:eq} is reached: 
$\omega=(_2^3)=3$. 
\end{quote}

\begin{figure}[ht]
\centerline{
\mbox{\epsfxsize=0.3\textwidth
    \epsfbox{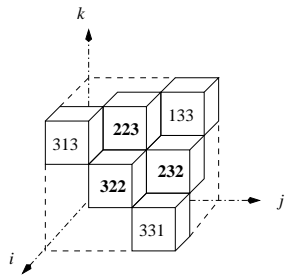}}}
\caption{Non generic example of ternary cubic, proved to be of maximal
rank: position of non-zero entries}\label{tensor333:fig}
\end{figure}

\begin{quote}
{\em Example.} Now take $\nbc=3$ and $d=3$. We are thus handling
$3\times3\times3$ symmetric tensors, or equivalently, ternary cubics.
The generic rank is 4, but the maximal rank is 5, according to table
\ref{Terncubics:table}. The class of maximal rank is unique, and a
representative is depicted in figure \ref{tensor333:fig}; the 6 non-zero
entries are all equal. Note that other non generic classes occur with
also a rank of 4, as pointed out in table \ref{Terncubics:table}.
\end{quote}

\begin{table}[!ht]
\begin{center}\begin{tabular}{l|l}
\hline
$\mathcal{GI}-$orbit & $\omega(p)$\\
\hline\hline
$x^3$ & 1\\
{\color{blue}$x^3+y^3$} & 2\\
$x^2 y$ & 3\\
$x^3+3\,y^2z$ & 4\\
$x^3 + y^3 + 6\,xyz$ & 4\\
$x^3 + 6\,xyz$ & 4\\
$a\,(x^3 +y^3 +z^3)+ 6b\,xyz$ & 4 ~({\bf generic})\\
$x^2y+xz^2$   & 5 \\
\hline
\end{tabular}\end{center}
\caption{Equivalence classes for ternary cubics: orbits under the action
of $\mathcal{GI}$, 
the group of invertible $3-$dimensional changes of
coordinates.}\label{Terncubics:table}
\end{table}

\begin{quote}
{\em Example.} Finally, consider ternary quartics, \ie, $(\nbc,d)=(3,4)$. 
In this case, the number of free parameters in the tensor is
$S=(^6_4)=15$. The number of free parameters in CAND exceeds 15 as soon as
$\omega\geq5$. So we could hope that we are lucky, because the number of
free parameters is the same on both sides of CAND. Unfortunately, this is
not the case, and Clebsh showed that
the generic rank was 6 \cite{EhreR93} (Ehrenborg and Rota
1993), as reported in table \ref{rank:table}. 
\end{quote}

%%%%%%%%%%%%%%%%%%%%%
\subsection{Rank$-1$ approximation}\label{approx:sec}
Approximating a tensor by another of rank 1 has at least two
applications in the present context. The first one is encountered when
when $\nbs\leq\nbc$ and when the source extraction  is performed one
source at a time in model \qref{model:eq}, contrary to section
\ref{diag:sec}; this is referred to as a {\em deflation} procedure. 

The maximization of the contrast \qref{maxCont:eq} then reduces to that of a
single output standardized cumulant (here the {\em kurtosis}), because a
single unit-norm vector is sought, instead of a whole orthogonal matrix:
\begin{equation}\label{maxKurt:eq}
\bw = Arg \Max_{||\bw||=1} \sum_{ijk\ell} w_i w_j w_k w_\ell
\,\cC^{\bms{y}}_{ijk\ell} 
\end{equation}
Yet, it has been shown \cite{DelaCD95:nolta} \cite{Como98:spie}
\cite{KofiR00:ams} (DeLathauwer Comon and others 1995; Comon 1998;
Kofidis and regalia 2000) that this maximization problem is equivalent
to minimizing $||\cC^{\bms{y}}-\sigma\,\bw\out\bw\out\bw\out\bw ||$,
which is simply finding the best rank$-1$ approximate of tensor
$\cC^{\bms{y}}$.

\smallskip

The second application is found in analytical BSS when sources are
discrete \cite{GrelC00:tampere} (Grellier and Comon 2000) or of
constant modulus \cite{VandP96:ieeesp} (Van der Veen 1996). In this
problem, we have to solve a system of $\nbe$ equations of the form
$(\bff\t\by_n)^d=1$, $1\leq n\leq\nbe$. This is equivalent to solving a
larger linear system $\bY\,\bff^{\krons d} =\un$, under the constraint
of $\Unvecs{}{\bff^{\krons d}}$ being a rank$-1$ tensor. Denote
$\{\bu_q^{\krons d}, 1\leq q\leq \bar{P}\}$ a basis of $\Ker{\bY}$. The
solution to this system takes the form
$$
\bff^{\krons d}=\bff_{min}^{\krons d} + \sum_{p=1}^{\bar{P}} \lambda_p\,
\bu_p^{\krons d} 
$$
where $\bff_{min}^{\krons d}$ is the minimum norm solution. Unfolding these
vectors in tensor form leads to the relation 
\begin{equation}\label{rank1comb:eq}
\bF = \bF_{min} +
\sum_{p=1}^{\bar{P}} \lambda_p\, \bU_p
\end{equation}
This problem can be shown to be related to the rank$-1$ combination
problem that we describe below.

%%%%%%%%%%%%%%%%%%%%%
\subsection{Rank$-1$ combination}\label{combin:sec}

{\bf The rank$-1$ combination problem} consists of finding the
numbers $\lambda_p$ so that, given matrices $\bU_p$, matrix $\sum_p
\lambda_p \bU_p$ has a rank of 1. Up to now, this problem has spawned
solutions that are not entirely satisfactory. As a consequence, so are
the solutions to \qref{rank1comb:eq}.

\medskip

Incidentally, we can restate the Joint
Approximate Diagonalization (JAD) problem addressed in
\cite{CardS93:ieeF} (Cardoso and Souloumiac 1993) for
the BSS into rank$-1$ combinations.

{\bf The Joint Approximate Diagonalization} of $\bar{P}$ matrices
$\bN_p$ consists of finding a square matrix $\bT$ such that $\bN_p
\approx \bT \bLambda_p \bT\t$, for all $p$, where $\bLambda_p$ are
diagonal matrices. From a property recalled in section
\ref{notation:sec}, this relation can be rewritten in vector form as
$\vecs{\bN_p}\eqdef \bn_p \approx \sum_i \lambda_{pi} \bt_i^{\krons2}$,
$\bt_i$ denoting the $i$th column of $\bT$. If the matrix
$[\lambda_{pi}]$ is full rank and has more columns than rows, then there
exists a matrix $\bB$ such that $\bt_j^{\krons2} \approx \sum_p B_{pj}
\bn_p$. Thus, given matrices $\bN_p$, the problem is to find
for every $j$, scalar coefficients $\beta_p$ such that $\sum_p \beta_p
\bN_p$ is a rank$-1$ matrix, and hence the link with the rank$-1$
combination problem.

\smallskip

However, the two problems are not equivalent, for
matrix $\bB$ is not necessarily square.

%%%%%%%%%%%%%%%%%%%%%
%%%%%%%%%%%%%%%%%%%%%
\section{Numerical algorithms}\label{algo:sec}
%%%%%%%%%%%%%%%%%%%%%
\subsection{Contrast maximization}
The ICA diagonalization of section \ref{diag:sec} (as well as the JAD
briefly mentioned in section \ref{combin:sec}) can be solved entirely
analytically in dimension $\nbc=2$, in a number of instances. In order
to exploit this property, Comon \cite{Como91:cha} \cite{Como94:SP} (1991)
proposed a sweeping of the pairs of indices, in a similar manner as in
the Jacobi diagonalization algorithm for Hermitian matrices
\cite{GoluV89} (Golub and Van Loan 1989). This idea has been later applied to JAD by Cardoso
\cite{CardS93:ieeF} (Cardoso and Souloumiac 1993).
To see this more in detail, consider the Givens rotation 
$$
\bQ = \left(\begin{array}{cc}
\cos\phi & \sin\phi\,\exp(\j \theta)\\
-\sin\phi\,\exp(-\j \theta) & \cos\phi
\end{array}\right)
$$
where the angle $\phi$ is imposed to lie in the interval $(-\pi/2,\:
\pi/2]$, because of inherent $\bLambda\bP-$indeterminacies.
Thus this matrix is entirely defined by the vector $\bu=[\cos2\phi,\:
\sin2\phi\,\cos\theta,\: \sin2\phi\,\sin\theta]$. 
Now, as in \qref{maxCont:eq}, define the contrast
$\Cont_{\alpha,d}$ as the sum of the $d-$th order tensor diagonal
entries raised to the power $\alpha$. Then it can be shown that 
$\Cont_{1,3}$ and $\Cont_{1,4}$ are real quadratic forms in $\bu$, and can thus
be easily maximized with respect to $\bu$, and hence to $(\theta,\phi)$
(by convention, if $\alpha=1$, the absolute value is dropped in
\qref{maxCont:eq}). On the other hand, this holds true for $\Cont_{2,3}$
but not for $\Cont_{2,4}$, which can be shown to be a quartic
\cite{Como94:SP} \cite{Como01:agadir} (Comon 1994; Comon 2001).
Nevertheless, polynomials of degree 4 can still be rooted analytically.

The procedure originally proposed by Comon (1989) consisting of sweeping
all the pairs, like in some numerical algorithms dedicated to matrices,
has never been proved to always lead to one of the
$\bLambda\bP-$equivalent absolute maxima, even if this is always
observed in practice. Counter-examples have never been found either. So
we consider this convergence issue as an open problem, belonging to the
general class of optimization problems over multiplicative groups.
However, some elements of convergence are now reported below.

\medskip

\paragraph{Convergence.}
For compactness, denote $\bG$ the cumulant tensor of the standardized
observation, $\bar{\by}$, which has been denoted $\cC^{\bar{\bms{y}}}$
up to now. Also 
denote $\bZ=\cC^{\bms{z}}$ the cumulant tensor obtained after an orthogonal
transformation $\bQ$. According to
the multi-linearity property, we have that:
\begin{equation}
Z_{pq..r} = \sum_{ij..\ell} Q_{pi} Q_{qj} \dots Q_{r\ell} \, G_{ij..\ell}
\end{equation}
Consider first the matrix case (order 2) in order to fix the ideas. The
contrast \qref{maxCont:eq} can then be written as:
\begin{equation}
\Cont_{2,2} = \sum_p |Z_{pp}|^2
\end{equation}
Because $\bQ$ is orthogonal, its differential can be written as
\begin{equation}
d\bQ = d\bS \, \bQ
\end{equation}
where matrix $\bS$ is skew-symmetric. This yields the relation
characterizing stationary points, $\bZ$: $\frac{1}{2}\,d\Cont_{2,2}=
2\sum_{p,t} Z_{pp} S_{pt} Z_{tp} = 0$. Yet, this is true for any skew-symmetric
matrix $\bS$, and hence for every skew-symmetric matrix having only two
non zero entries (one $+1$ and one $-1$); based
on this argument, one concludes that:
\begin{equation}\label{station2:eq}
(Z_{qq}-Z_{rr}) Z_{qr} = 0, \: {\rm for} \; q\neq r
\end{equation}
Next, the local convexity can be examined with the help of the same
tools, observing that:
\begin{equation}\label{convex2:eq}
\frac{1}{4}d^2\Cont_{2,2}=4 Z_{qr}^2 -
(Z_{qq}-Z_{rr})^2
\end{equation}
Thus, there are three kinds of stationary points:
(i)~those for which all 
diagonal entries are equal, which correspond to minima of $\Cont_{2,2}$,
(ii)~those for which all non-diagonal entries are null, which correspond to
maxima, and (iii)~saddle points, for which some diagonal entries are
equal and some non-diagonal entries vanish. This result is well known,
and proves that the only maxima are diagonal matrices, which can be
deduced from each other by mere permutation within the diagonal.

Now let us develop the same calculations for tensors of order 3 and 4.
Stationary values are given by the relations:
\begin{eqnarray*}
\frac{1}{2}\,d\Cont_{2,3} &=& 3\sum_{p,t} Z_{ppp}\,dS_{pt}\, Z_{tpp}=0, \\
\frac{1}{2}\,d\Cont_{2,4} &=& 4\sum_{p,t} Z_{pppp}\,dS_{pt}\, Z_{tppp}=0
\end{eqnarray*}
or, on the basis of skew-symmetric matrices, for $q\neq r$:
\begin{eqnarray}
Z_{qqq} Z_{qqr} - Z_{rrr} Z_{qrr} & = & 0, \label{station3:eq}\\
Z_{qqqq} Z_{qqqr} - Z_{rrrr} Z_{qrrr} & = & 0,\label{station4:eq}
\end{eqnarray}
whereas local convexity conditions are governed by (Comon 1994):
\begin{eqnarray}
\frac{1}{6}d^2\Cont_3\!\!&{\!\!=\!\!}&\!\!4Z_{qqr}^2 +\!4Z_{qrr}^2
\!-(Z_{qqq}\!-Z_{qrr})^2\! -(Z_{rrr}\!-Z_{qqr})^2 \label{convex3:eq}\\
\frac{1}{8}d^2\Cont_4\!\!&{\!\!=\!\!}&\!\!{\frac{9}{2}}Z_{qqrr}^2\! +
4Z_{qqqr}^2\! + 4Z_{qrrr}^2 
\!-(Z_{qqqq}-{\frac{3}{2}}Z_{qqrr})^2 \!-
(Z_{rrrr}\!-{\frac{3}{2}}Z_{qqrr})^2 ~~\quad~~ \label{convex4:eq}
\end{eqnarray}
The comparison of these results with \qref{station2:eq} and
\qref{convex2:eq} lead to two conclusions: (a)~non-diagonal terms do not
factorize anymore in \qref{station3:eq} and \qref{station4:eq}, so that
stationary values are more difficult to characterize, and (b)~diagonal
tensors are still local maxima, but there are {\em a priori} others.
This is another problem, linked to optimization in groups, that this
author considers as open. 

\paragraph{Sweeping strategies.}
We have presented several numerical algorithms aiming at separating
$\nbs=2$ sources 
from $\nbc=2$ sensors in the presence of noise of unknown statistics.
Inspired from the Jacobi cyclic-by rows sweeping strategy proposed for
matrices, we can process all the $\nbc(\nbc-1)/2$ pairs one by one
sequentially (Comon 1989; Comon 1994). However, as in the matrix case,
the noise part (constituted by the actual background noise and all the
other $\nbc-2$ sources) changes at every step, so that a single sweeping
is not sufficient. In practice, an order of $\sqrt{\nbc}$ sweeps have
been shown to be sufficient.

Other strategies have been analyzed, and consist of processing first the
pair of sensors that yields the maximal increase in the contrast
criterion. This strategy has also been implemented successfully, but is
not always numerically efficient.

When processing one pair $(i,j)$, one can either recompute all the entries of
the cumulant tensor that have been affected (\ie, those whose indices
contain $i$ or $j$), or compute the rotated data instead. The two
possibilities do not have the same numerical complexity, and the best
choice depends on the number of sensors, $\nbc$, and on the number of
samples, $\nbe$.

%%%%%%%%%%%%%%%%%%%%%
\subsection{Parafac algorithm}
In \cite{LeurRA93:simax} (Leurgans et alterae 1993), SVD-based
algorithms are proposed to compute CAND of $3$rd order tensors in larger
dimensions. However, these algorithms, called {\sc Parafac}, need the
number of sources, $\nbs$, to be smaller than or equal to
$\frac{3}{2}\,\nbc-1$, in the symmetric case we are interested in. See
also \cite{Krus77:laa} \cite{Bro97:cils} (Kruskal 1977; Bro 1997)
for more details. In view of table \ref{rank:table} reported above,
this value of $\nbs$ is strictly smaller than the generic rank,
$\omega$, except for $(d,\nbc)=(3,2)$ or $(d,\nbc)=(3,4)$. As a
consequence, {\sc Parafac} algorithms can only {\em approximate} $d-$way
arrays, in general.

In the unsymmetric problem, the goal is to find three matrices, $\bA$,
$\bB$, and $\bC$, such that $G_{ijk} = \sum_p A_{ip} B_{jp} C_{kp}$.
One possible numerical algorithm is based on alternating least squares,
as explained below for $3-$way arrays \cite{CarrC70:psy} (Carroll and
Chang 1970): 
\begin{itemize}
\item
Start with ($\bA(0)$, $\bB(0)$, $\bC(0)$)
\item
Define matrices $\bG^{(1)}$, $\bG^{(2)}$, $\bG^{(3)}$:\\
$\bG_{ijk} =\bG^{(1)}_{ip}=\bG^{(2)}_{jq}=\bG^{(3)}_{kr}$;~
$p=(jk)$, $q=(ik)$, $r=(ij)$
\item
Estimate stage $t+1$ from stage $t$ by pseudo-inversion:
\begin{itemize}
\item
Update mode 1:\\
$\bA(t+1)=\bG^{(1)}\,[\bB(t)\t \: \bC(t)\t]\pinv$ 
\item
Update mode 2:\\
$\bB(t+1)=\bG^{(2)}\,[\bA(t+1)\t \: \bC(t)\t]\pinv$ 
\item
Update mode 3:\\
$\bC(t+1)=\bG^{(3)}\,[\bA(t+1)\t \: \bB(t+1)\t]\pinv$ 
\end{itemize}\end{itemize}
where $\bM\pinv$ denotes the Moore-Penrose pseudo inverse of $\bM$. See
also \cite{Bro97:cils} \cite{Dela97:these} \cite{Krus77:laa} (Bro
1997; DeLathauwer 1997; Kruskal 1977) for more details on {\sc Parafac}
algorithms.

%%%%%%%%%%%%%%%%%%%%%
\subsection{Sylvester theorem}\label{sec:sylvester}
As already pointed out earlier, a rank-one tensor is associated with a
linear form raised to the $d$th power. In terms of polynomials, the CAND
decomposition can thus be rephrased: how can one decompose a quantic
into a sum of $d$th powers of linear forms \cite{ComoM96:SP} (Comon
and Mourrain 1996)~? This is this topic that addresses this theorem,
restricted to the binary case however (\ie, two variables).

\begin{theorem}
A binary quantic $p(x,y) = \sum_{i=0}^d \gamma_i \, c(i) \,
x^i \, y^{d-i}$ can be written as a sum of $d$th
powers of $\omega$ distinct linear forms:
$$
p(x,y) = \sum_{j=1}^\omega \lambda_j \: (\alpha_j \, x + \beta_j \, y)^d,
$$
if and only if (i)~there exists a vector $\bg$ of dimension $\omega+1$, with
components $g_\ell$, such that
\begin{equation}\label{eq:sylves}
\left[\begin{array}{cccc}
\gamma_0 & \gamma_1 & \cdots & \gamma_\omega\\
\vdots   &  &    &  \vdots\\
\gamma_{d-\omega} & \cdots & \gamma_{d-1} & \gamma_d
\end{array}\right] \: \bg = \zero.
\end{equation}
and (ii)~the polynomial $q(x,y) \eqdef \sum_{\ell=0}^\omega g_\ell\,
x^\ell\, y^{\omega-\ell}$ admits $\omega$ distinct roots.
\end{theorem}

\smallskip

Sylvester's theorem not only proves the existence of the $\omega$ forms
(second column in the tables), but also gives a means to compute them
\cite{Como98:spie} \cite{ComoM96:SP} (Comon 1998; Comon and Mourrain
1996). For odd values of $d$, we have thus a generic rank of
$\omega=\frac{d+1}{2}$, whereas for even values of $d$,
$\omega=\frac{d}{2}+1$. So when $d$ is odd, there is generically a
unique vector $\bg$ satisfying \qref{eq:sylves}, but there are two of
them when $d$ is even. This theorem shows that in column $\nbc=2$ of
table \ref{dimD:table}, we have $D=0$ when $d$ is odd, and $D=1$ when $d$
is even.

In \cite{DelaCD99:caesarea} (DeLathauwer Comon and DeMoor 1999),
several extensions to this theorem are proposed in the complex case. The
basic idea remains the same, but the result becomes more complicated.

The disappointing fact is that Sylvester's theorem cannot be extended to
dimensions higher than 2. In fact, a key step in the proof
\cite{ComoM96:SP} \cite{Como98:spie} (Comon and Mourrain 1996; Comon
1998) is that for any polynomial $p$ of degree $d$, and any monomial $m$
of degree $d-\omega$, there exists a polynomial $q$ of degree $\omega$
such that $qm$ is orthogonal to $p$. Equation \qref{eq:sylves} expresses
that orthogonality in terms of polynomial coefficients. It is clear that
this holds true only when $d\geq\omega$, which is unfortunately
satisfied only in the binary case, according to table \ref{rank:table}.
Possibilities of extension to more than 2 variables is discussed in
\cite{ComoM96:SP} (Comon and Mourrain 1996).

%The computation of CAND (section \ref{cand:sec}) has been carried 
%out only in the restrictive case of dimension $\nbc=2$.

\paragraph{Simultaneous CAND. }
Let us go back to table \ref{dimD:table}. Among others, this table
reports that there are infinitely many CAND for even orders, $d$.
In order to fix this indeterminacy in the case $(d,\nbc,\nbs)=(4,2,3)$ (the
manifold of solutions is of dimension 1 in that situation), it is proposed in
\cite{Como98:spie} (Comon 1998) to
simultaneously diagonalize a second cumulant tensor of order 4.

\paragraph{The help of virtual sources. }
In \cite{Como98:spie} \cite{ComoG99:aussois} (Comon 1998; Comon and
Grellier 1999) an algorithm dedicated to discrete 
sources is proposed, and performs both the identification of $\bA$ and
the extraction of sources $x_i$, in the case $(d,\nbc,\nbs)=(2,2,3)$.

In a few words, assume three sources $x_i$ are mixed and received on two
sensors, and assume these sources are all distributed in $\{-1,\,+1\}$
(they are called BPSK in digital communications). One can prove, if
sources $x_i$ are statistically independent, that the ``virtual'' source
$x_1x_2x_3$ is also BPSK-distributed, but obviously statistically
dependent of the three former ones. However, one can still prove that all its
fourth-order pairwise cross-cumulants vanish. Yet, only {\em pairwise}
cumulants are utilized in the sweeping strategies maximizing contrasts
such as $\Cont_{2,4}$ in \qref{maxCont:eq}. As a consequence, viewed by
the algorithm, sources are independent; one can thus build
from $\by\t=[y_1,y_2]$ virtual measurements $y_1^3$, $y_1^2y_2$,
$y_1y_2^2$, and $y_2^3$, that can be modeled as linear mixtures of 4th
order pairwise independent unknown sources. This allows the separation
of the four sources (three actual and one virtual) from six sensors (two
actual and four virtual).

%%%%%%%%%%%
\subsection{Rank-one approximation}

The rank$-1$ approximation problem (section \ref{approx:sec}) has been
partly solved by algorithms inspired from the matrix {\em power method}
and devised for arrays of higher orders \cite{DelaCD95:nolta}
\cite{Dela97:these} \cite{KofiR00:ams} (DeLathauwer Comon and others
1995; DeLathauwer 1997; Kofidis and Regalia 2000).
\smallskip

\paragraph{Criteria.}
Given tensor $\cC^{\bms{y}}$, the goal is to find a vector $\bw$ minimizing:
\begin{equation}\label{omega0:eq}
\Omega_o = ||\cC^{\bms{y}}-\sigma\,\bw\out\bw\out\bw\out\bw||
\end{equation}
for some scalar number $\sigma$.
One can prove that minimizing \qref{omega0:eq} is equivalent to
maximizing
\cite{Como98:spie} (Comon 1998):
\begin{equation}
\Omega_d = ||\cC^{\bms{y}}\con\bw\con\bw\con\dots\con\bw||
\end{equation}
or to minimizing:
\begin{equation}
\Omega_{d-1} = ||\cC^{\bms{y}}\con\bw\con\dots\con\bw-\lambda\,\bw||
\end{equation}
However, the other criteria $\Omega_r$, $0<r<d-1$, are generally not
equivalent. 

\paragraph{Stationary uplets} $\!(\bv,\lambda)$ of $\Omega_o$,
$\Omega_{d-1}$ or $\Omega_d$ are the same and satisfy:
$$
\cC^{\bms{y}}\con\underbrace{\bv\con\dots\con\bv}_{d-1 ~ {\rm
times}}=\lambda\,\bv 
$$
this suggests a Rayleigh-like iteration, tat we can call the
{\em Tensor Rayleigh symmetric iteration}:
\begin{eqnarray*}
\bw &\leftarrow& \cC^{\bms{y}}\con\underbrace{\bw\con\dots\con\bw}_{d-1 ~ {\rm
times}}\\ 
\bw &\leftarrow& \bw/||\bw||
\end{eqnarray*}
In \cite{DelaCD95:nolta} (Delathauwer Comon et al. 1995), it is
suggested to run a non symmetric iteration, and to initialize the
algorithm with the HOSVD. 

\paragraph{The rank$-1$ combination problem} (section \ref{combin:sec})
has been solved in a sub-optimal way up to now in \cite{VandP96:ieeesp}
\cite{GrelC00:tampere} (Van der Veen and Paulraj 1996; Grellier and
Comon 2000) by solving a large unconstrained linear system,
and trying to restore the structure afterwards. The optimal
one-stage solving still remains to be devised.

%%%%%%%%%%%%%%%%%%%%%
%%%%%%%%%%%%%%%%%%%%%
\section{Concluding remarks}
In this chapter, we have partly surveyed the tools dedicated to tensor
decompositions, mainly through the problem of source separation. Thus,
this presentation has been restrictive, but hopefully still
informative.

Many other source separation algorithms do not resort to tensor tools,
and have not been reported here. It is worth noting that some of them do
not need the sources to be statistically independent, so that the output
cumulant tensor is not even aimed at being diagonal. Instead, other
properties of the sources can be exploited, such as their discrete
character, or their constant modulus \cite{VandP96:ieeesp}
\cite{TalwVP96:ieeesp} \cite{GrelC00:tampere} (Van der Veen and
Paulraj 1996; Talwar Viberg and Paulraj 1996; Grellier and Comon 2000).
When more sources than sensors are present, general results state that
it is sometimes possible to identify the mixture, but source extraction
requires more knowledge about the sources (\eg, their distribution).
These issues have been tackled herein. Let us now turn to research
perspectives.

\medskip

In the area of source separation, current hot research topics include
(i)~blind identification of under-determined mixtures, (ii)~blind
equalization of convolutive mixtures, (iii)~the theoretical proof of
convergence of pair-sweeping algorithms, and, in the context of
telecommunications, (iv)~handling properly carrier residuals when
present in the measurements. In all cases, analytical block-algorithms
are suitable when computer power is available and when the stationarity
duration is short.

As far as tensors are concerned, open research directions include:
(i)~the determination of the maximal achievable rank for arbitrary order
and dimensions, (ii)~the actual calculation of general Canonical
Decompositions for $\nbc>2$, (iii)~efficient numerical algorithms for
computing an approximate of given rank.

\end{document}